\documentclass[prd,preprint,superscriptaddress,showpacs]{revtex4}
\usepackage{revsymb}
\usepackage[dvips]{graphicx}

\usepackage{epsfig}
\usepackage{graphicx}
\usepackage{indentfirst}
\usepackage{amsmath}
\usepackage{amsfonts}
\usepackage{amssymb}

\begin{document}
\title{Transient cosmic acceleration from interacting fluids}

\author{J\'ulio C. Fabris}
\email{fabris@pq.cnpq.br}
\affiliation{Universidade Federal do Esp\'{\i}rito Santo,
Departamento
de F\'{\i}sica\\
Av. Fernando Ferrari, 514, Campus de Goiabeiras, CEP 29075-910,
Vit\'oria, Esp\'{\i}rito Santo, Brazil}

\author{Bernardo Fraga}
\email{bernardo@cbpf.br}
\affiliation{ICRA - Centro Brasileiro de
Pesquisas F\'{\i}sicas -- CBPF, \\ Rua Xavier Sigaud, 150, Urca,
CEP 22290-180, Rio de Janeiro, Brazil}

\author{Nelson Pinto-Neto}
\email{nelson.pinto@pq.cnpq.br}
\affiliation{ICRA - Centro Brasileiro de
Pesquisas F\'{\i}sicas -- CBPF, \\ Rua Xavier Sigaud, 150, Urca,
CEP 22290-180, Rio de Janeiro, Brazil}

\author{Winfried Zimdahl}
\email{winfried.zimdahl@pq.cnpq.br}
\affiliation{Universidade Federal do Esp\'{\i}rito Santo,
Departamento
de F\'{\i}sica\\
Av. Fernando Ferrari, 514, Campus de Goiabeiras, CEP 29075-910,
Vit\'oria, Esp\'{\i}rito Santo, Brazil}

\date{\today}

\begin{abstract}
Recent investigations seem to favor a cosmological dynamics according to which the accelerated expansion of the Universe may have already peaked and is now slowing down again \cite{sastaro}. As a consequence, the cosmic acceleration may be a transient phenomenon. We investigate a toy model that reproduces such a  background behavior  as the result of a time-dependent coupling in the dark sector which implies a cancelation of the ``bare" cosmological constant.
With the help of a statistical analysis of Supernova Type Ia (SNIa) data we demonstrate that for a certain parameter combination a transient accelerating phase emerges as a pure interaction effect.
\end{abstract}

\maketitle

\section{Introduction}

By now there is large direct and indirect evidence that the  Universe entered a period of accelerated expansion. Direct evidence is provided by the luminosity-distance data of
supernovae of type Ia \cite{SNIa} (see, however, \cite{sarkar}), indirect evidence comes from the anisotropy spectrum of the cosmic microwave background radiation (CMBR) \cite{cmb}, from large-scale-structure data \cite{lss}, from the integrated Sachs--Wolfe effect
\cite{isw}, from baryonic acoustic
oscillations \cite{eisenstein} and from gravitational lensing \cite{weakl}.
According to the currently prevailing interpretation, based on Einstein's GR,  our Universe is dynamically dominated by two so far unknown components, dark matter (DM) and dark energy (DE). The latter contributes roughly 75\% to the total energy budget, the former about 20\%. Only about 5\% are in the form of conventional, baryonic matter. The preferred model is the $\Lambda$CDM model which also plays the role of a reference model for alternative approaches to the DE problem. While the $\Lambda$CDM model can describe most of the observations, there still remain puzzles \cite{peri}. The present situation in the field is characterized, e.g., in \cite{rev,pad,dumarev} and references therein.

Both DM and DE manifest themselves observationally only through their gravitational action.
In most studies both these components are regarded as independent.
However, there exists a line of investigation that explores the consequences of a coupling within the dark sector. Corresponding models show a richer dynamics than models with separately conserved energy-momentum tensors.
It has been argued that ignoring a potentially existing interaction between both components may give rise to misinterpretations of observational data regarding the equation-of-state (EoS) parameter \cite{Das}.
A coupling between DM and DE may also be relevant with respect to the coincidence problem (see, e.g. \cite{coinc} and references therein).
Models with an interaction matter--dark
energy were introduced by Wetterich \cite{wetterich}. Meanwhile there exists a still growing body of
literature on the subject -see, e.g. \cite{interacting} and
references therein.
Limits for the admissible interaction strength have been obtained for various configurations \cite{limits}.
On the other hand, the transition from decelerated to accelerated expansion can be understood as a pure interaction phenomenon in the context of holographic DE models \cite{DW,essay,WDCQG}.

Recently it was argued that the latest SNIa could favor a scenario in which the cosmic acceleration has past a maximum value and is now slowing down again \cite{sastaro}. In such a case the accelerated expansion could be a transient phenomenon with a late-time dynamics, incompatible with that of the $\Lambda$CDM model. It is this possibility which we are going to study in the present paper.
We mention that models of transient acceleration were already discussed in \cite{alcaniz} and more recently in \cite{alcaniztr}.

It is the purpose of this paper to provide a toy model which describes a transient cosmological acceleration as the consequence of an interaction between dark matter and dark energy. Such a dynamics cannot be obtained if the interaction represents a small correction to, say, the $\Lambda$CDM model. For models of this type  the long-time cosmological dynamics will always be determined by the cosmological term and result
in accelerated expansion. To achieve transient accelerated expansion, a twofold role of the interaction is necessary. At first, it has to cancel the ``bare" cosmological constant and at second it has to generate a phase of accelerated expansion by itself. Acceleration has to be an interaction phenomenon.
We show that these requirements can be fulfilled by interaction terms that combine powers and exponentials of the cosmic scale factor.

The paper is organized as follows. In Sec. \ref{Interacting fluid dynamics} we introduce the basic interacting fluid model and study the influence of the coupling, described by an initially arbitrary function of the scale factor, on the cosmic acceleration equation.  Sec. \ref{transient} investigates two interaction types between DM and DE, each of them characterized by two parameters, in detail. Conditions for the existence of transient acceleration are found under the condition of a vanishing total effective cosmological constant.
In Sec. \ref{SN} we use a statistical analysis of the gold sample of the SNIa data \cite{gold} to find the admissible parameter ranges for transient cosmic acceleration. The best-fit models for both interactions
are shown to be compatible with the $\Lambda$CDM model. Our conclusions are summarized in Sec. \ref{conclusions}.

\section{Interacting fluid dynamics}
\label{Interacting fluid dynamics}

We model the present cosmic substratum as a mixture of three components: dark energy (DE) with an energy density $\rho_x$, dark matter (DM) with an energy density $\rho_{m}$ and baryonic matter with an energy density $\rho_{b}$.
The Friedmann equation in a spatially flat universe is
\begin{equation}
3 H^{2} = 8\,\pi\,G\,\left(\rho_x + \rho_m + \rho_b\right)\ ,
 \,  \label{fried}
\end{equation}
where $H=\dot{a}/a$ is the Hubble parameter and $a$ is the scale factor of the Robertson-Walker metric. The DE component is supposed to be characterized by
an equation of state $p_x = w \rho_x$. We admit an interaction between DE and DM according to
\begin{eqnarray}
\dot{\rho_x}+3H(1+w)\rho_x &=&-Q\ , \nonumber \\
\dot{\rho}_{m}+3H\rho_{m}&=&Q \ , \label{interacao}
\end{eqnarray}
where $Q$ symbolizes the interaction, and a
dot represents a derivative with respect to the cosmic time.
The baryon component is separately conserved,
\begin{equation}
\dot{\rho}_{b}+3H\rho_{b}=0 \quad \Rightarrow\quad \rho_{b} = \rho_{b_{0}}\left(\frac{a_{0}}{a}\right)^{3}\ .
\label{ba}
\end{equation}
The split of the total energy density
$\rho$ into $\rho = \rho_{m} + \rho_{b} + \rho_{x}$ is consistent with the energy conservation for $\rho$.
We assume, without loss of generality, that the energy density of the dust fluid
can be written as:
\begin{equation}
\rho_m=\tilde\rho_{m_0}\left(\frac{a_0}{a}\right)^{3}\,f\left(a\right)\ ,
\label{materia}
\end{equation}
where $\tilde\rho_{m_0}$ and $a_0$ are constants, and $f(a)$ is an
arbitrary time-dependent function. It follows from (\ref{interacao})
and (\ref{materia}) that
\begin{equation}\label{Q}
Q=\rho_m\frac{\dot{f}}{f}=\tilde\rho_{m_0}\left( \frac{a_0}{a} \right)^3\dot{f} \ .
\end{equation}
Let us write $f(a)$ as
\begin{equation}
\label{f}
f\left(a\right)=1+g\left(a\right)\ ,
\end{equation}
such, that the deviation from the non-interacting case is encoded in the function $g\left(a\right)$.
With
\begin{equation}
\dot{f}=\dot{g} = \frac{d g}{d a}\dot{a}\
\label{df}
\end{equation}
we obtain
\begin{equation}\label{Q2}
Q=\tilde\rho_{m_0}\frac{d g}{da}\dot{a}\left(\frac{a_0}{a} \right)^3 .
\end{equation}
For $\rho_{m}$ it follows that
\begin{equation}
\rho_m=\tilde\rho_{m_0}\left(1+g\right)\left(\frac{a_0}{a}\right)^{3}\ .
\label{rm}
\end{equation}
Denoting the value of $\rho_m$ at $a= a_{0}$ by
${\rho}_{m_{0}}$, both initial values are related by
\begin{equation}
{\rho}_{m_{0}} = \tilde\rho_{m_{0}}\left(1 + g_{0}\right) \ ,
\label{tilm0}
\end{equation}
where $g_{0} \equiv g(a_{0})$. The quantity
${\rho}_{m_{0}}$ is the value of $\rho_{m}$ at $a=a_{0}$ in
the presence of the interaction, $\tilde\rho_{m_{0}}$ is the value of
$\rho_{m}$ at $a=a_{0}$ for vanishing interaction. The interaction
re-normalizes the present value of $\rho_{m}$.
According to (\ref{Q2}), a positive value of $Q$, i.e., an energy transfer from dark energy to dark matter requires
$\frac{d g}{d a} >0$ in an expanding universe. For $\frac{d g}{d a} <0$ the transfer is in the opposite direction, i.e., $Q<0$.

With (\ref{Q2}), the energy density of the $x$-component is determined by
\begin{equation}
\dot{\rho}_x+3H(1+w)\rho_x =-\tilde\rho_{m_0}\frac{d g}{da}\dot{a}\left(\frac{a_0}{a} \right)^3  \ .
\label{drx}
\end{equation}
For a constant EoS parameter $w$ the solution of (\ref{drx}) is
\begin{equation}
\rho_x = \rho_{x_{0}}\left(\frac{a_0}{a}\right)^{3\left(1+w\right)} - \tilde\rho_{m_0}a_{0}^{3}\,a^{-3\left(1+w\right)}\, \int_{a_{0}}^{a} da
\frac{d g}{d a}\,a^{3w}\ .
\label{rx3}
\end{equation}
Integrating by parts yields
\begin{equation}
\rho_x = \left(\rho_{x_{0}} + \tilde\rho_{m_0} g_{0}\right)\left(\frac{a_0}{a}\right)^{3\left(1+w\right)} - \tilde\rho_{m_0}\left(\frac{a_0}{a}\right)^{3}\,g
+  3w\,\tilde\rho_{m_0}a_{0}^{3}\,a^{-3\left(1+w\right)}\,\int_{a_{0}}^{a} da\,
g\,a^{3w - 1}\ .
\label{rx5}
\end{equation}
For a given interaction $g(a)$, the Hubble rate in Eq.~(\ref{fried}) is then determined by the sum of $\rho_{b}$ from (\ref{ba}), $\rho_{m}$ from (\ref{rm}) and $\rho_{b}$ from (\ref{rx5}).

The acceleration equation
\begin{equation}
\frac{\ddot{a}}{a} = - \frac{4\pi G}{3}\left[\rho_{m} + \rho_{b} + \left(1 + 3w\right)\rho_{x}\right]\
\label{acc}
\end{equation}
takes the form
\begin{eqnarray}
\frac{\ddot{a}}{a} &=&- \frac{4\pi G}{3}\left\{\tilde\rho_{m_0}\left(1+g\right)\left(\frac{a_0}{a}\right)^{3} + \rho_{b_{0}}\left(\frac{a_{0}}{a}\right)^{3} + \left(1 + 3w\right)\left[\left(\rho_{x_{0}} + \tilde\rho_{m_0} g_{0}\right)\left(\frac{a_0}{a}\right)^{3\left(1+w\right)}\right.\right. \nonumber\\
 && \left.\left. \qquad\qquad - \tilde\rho_{m_0}\left(\frac{a_0}{a}\right)^{3}\,g
+  3w\,\tilde\rho_{m_0}a_{0}^{3}\,a^{-3\left(1+w\right)}\,\int_{a_{0}}^{a} da\,
g\,a^{3w - 1}\right]\right\} \ .
\label{dda}
\end{eqnarray}
It can also be written as
\begin{eqnarray}
\frac{\ddot{a}}{a} &=&- \frac{1}{2}H_{0}^{2}\left\{\tilde\Omega_{m_0}\left(\frac{a_0}{a}\right)^{3} + \Omega_{b_{0}}\left(\frac{a_{0}}{a}\right)^{3} + \left(1 + 3w\right)\left[\left(\Omega_{x_{0}} + \tilde\Omega_{m_0} g_{0}\right)\left(\frac{a_0}{a}\right)^{3\left(1+w\right)}\right]\right. \nonumber\\
 && \left. \qquad\qquad + 3w\,\tilde\Omega_{m_0}\left(\frac{a_0}{a}\right)^{3}\,\left[\left(1 + 3w\right)
 a^{-3w}\,\int_{a_{0}}^{a} da\,g\,a^{3w - 1}  - g\right]\right\}
 \ ,
\label{q}
\end{eqnarray}
where we have introduced the parameters $\tilde\Omega_{m_0} = \frac{8\pi G \tilde\rho_{m_0}}{3 H_{0}^{2}}$, $\Omega_{m_0} = \frac{8\pi G \rho_{m_0}}{3 H_{0}^{2}}$,  $\Omega_{x_0} = \frac{8\pi G \rho_{x_0}}{3 H_{0}^{2}}$ and $\Omega_{b_0} = \frac{8\pi G \rho_{b_0}}{3 H_{0}^{2}}$.
For any specific model $g(a)$, formula (\ref{q}) describes the dynamics of $\frac{\ddot{a}}{a}$ through the cosmic history, as long as our three-component model makes sense.

The present value of the deceleration parameter is
\begin{equation}
\frac{\ddot{a}}{aH^{2}}|_{0} = - \frac{1}{2}\left\{\tilde\Omega_{m_0}\left(1 + g_{0}\right) + \Omega_{b_{0}} + \left(1 + 3w\right)\Omega_{x_{0}} \right\} = - \frac{1}{2}\left\{{\Omega}_{m_0} + \Omega_{b_{0}} + \left(1 + 3w\right)\Omega_{x_{0}} \right\}
 \ ,
\label{q01}
\end{equation}
where ${\Omega}_{m_0} = \tilde\Omega_{m_0}\left(1 + g_{0}\right)$. In a spatially flat universe with $\Omega_{m_0} + \Omega_{b_{0}} = 1 - \Omega_{x_0}$, Eq.~(\ref{q01}) reduces to
\begin{equation}
\frac{\ddot{a}}{aH^{2}}\mid _{0} =  - \frac{1}{2}\left\{1 + 3w\Omega_{x_{0}}\right\}
 \ .
\label{q0x}
\end{equation}
Notice that there is no \textit{direct} influence of the interaction on the present value of the deceleration parameter \cite{statef}.

\section{Transient accelerating phases in the Universe}
\label{transient}

\subsection{Specifying the interaction }

Lacking any knowledge about the nature of dark energy and that of dark matter, there is no well-motivated choice for the interaction either. Our strategy here is as follows. Since we try to model a transient accelerating universe, we postulate a mathematically tractable interaction which presumably is compatible with such a dynamics. We solve this dynamics and clarify whether there exist parameter combinations that are favorable for a scenario of transient acceleration. It will turn out that for the two-parameter interactions considered below, there are indeed such ranges.
Subsequently, we compare the result with the SNIa observational
data.
We shall consider couplings for which the function $g(a)$ is represented by a combination of powers and exponentials of the scale factor. In particular, we
investigate the following two types of interaction terms: $g_1(a)=c_1
a^n \exp (-a/\sigma)$ and $g_2(a)=c_2 a^n \exp (-a^2/\sigma ^2)$,
where $n$ is a positive integer and $\sigma$ is a positive real
number.
Although these choices may seem arbitrary, they will admit a suitable parameter combination and
they will provide us with a transparent picture concerning the potential role of interactions within the dark sector in a scenario of transient acceleration.

In the first case, i.e., for $g_1(a)=c_1
a^n \exp (-a/\sigma)$, assuming that $3w+n-1$ is a natural number, the energy density (\ref{rx5}) is given by
\begin{eqnarray}
\rho_{x} &=& \rho_{x_{0}}^{eff} \left(\frac{a_0}{a}\right)^{3\left(1 + w\right)} - K_1 a_{0}^{3}a^{n-3}\,\exp\left(-a/\sigma\right)\nonumber\\
&& \quad - 3w K_1 a^{-3\left(1+w\right)}a_{0}^{3}\,\exp\left(-a/\sigma\right)
\sum_{i=0}^{n+3w-1}\left[\frac{(n+3w-1)!}{i!}\sigma^{n+3w-i}a^{i}\right]\ ,
\label{rh1}
\end{eqnarray}
where
\begin{equation}
\rho_{x_{0}}^{eff} = \rho_{x_{0}} + K_{1}a_{0}^{n}\,\exp\left(-a_{0}/\sigma\right) + 3w K_1 \,\exp\left(-a_{0}/\sigma\right)
\sum_{i=0}^{n+3w-1}\left[\frac{(n+3w-1)!}{i!}\sigma^{n+3w-i}a_{0}^{i-3w}\right]\
\label{reff}
\end{equation}
and $K_1\equiv c_1\tilde\rho_{m_0}$.
From now on we shall put $a_{0} = 1$.

The expression (\ref{dda}) specifies to
\begin{eqnarray}
 \frac{\ddot a}{a}&=&-\frac{4\pi
G}{3}\left[\frac{\tilde\rho_{m_0}}{a^3} + \frac{\rho_{b_0}}{a^3}
+(1+3w)\frac{\rho_{x_0}^{eff}}{a^{3(1+w)}}\right]\nonumber
\\&+&4\pi G w K_1 \exp (-a/\sigma)
\left[(1+3w)\sum_{i=0}^{n+3w-1}\frac{(n+3w-1)!}{i!}\sigma^{n+3w-i}a^{i-3(1+w)}
+a^{n-3}\right]\ .\nonumber\\
\label{ddah1}
\end{eqnarray}

Let us now analyze the special case $w=-1$ and $n=5$ for which the sum in (\ref{rh1}) and (\ref{reff}) has only two terms.
Using $\tilde\rho_{m_0} = \rho_{m_0} - K_{1}\,\exp (-1/\sigma)$, the matter density (\ref{rm}) reduces to
\begin{equation}
\label{rhom1}
\rho_{m} = \rho_{m_0}a^{-3}  + K_{1}a^{-3}\left[a^{5}\exp(-a/\sigma) - \exp(-1/\sigma)\right]\ ,
\end{equation}
and the density (\ref{rh1}) of the $x$ fluid becomes
\begin{equation}
\rho_{x} = \rho_{x_{0}}^{eff}
+ 3K_{1}\,\exp\left(-a/\sigma\right) \left(\sigma^{2} + \sigma a - \frac{1}{3}a^{2}\right)
\ ,
\label{rh1a}
\end{equation}
with
\begin{equation}
\label{reffh12}
\rho_{x_{0}}^{eff} = \rho_{x_{0}} - 3K_{1}\,\exp(-1/\sigma)
\left[\sigma^{2} + \sigma - \frac{1}{3}\right]\ .
\end{equation}
For a vanishing interaction the $x$ component coincides with a cosmological constant.
For the acceleration equation (\ref{ddah1}) one finds
\begin{equation}
\label{accOmega1}
\frac{\ddot a}{a}=-\frac{1}{2}H_{0}^{2}
\left\{\frac{\Omega_{m_{0}} + \Omega_{b_0} - \bar{K}_{1}\exp (-1/\sigma)}{a^3}  -2\Omega_{x_0}^{eff} + 3 \bar{K}_1 \exp
(-a/\sigma) \left[a^{2} - 2\sigma^{2} - 2\sigma a\right]\right\}
\end{equation}
with $\bar{K}_{1} = \frac{8 \pi G}{3 H_{0}^{2}}K_{1}$ and  $\Omega_{x_0}^{eff} = \frac{8 \pi G}{3 H_{0}^{2}}\rho_{x_0}^{eff}$.

For an interaction  $g_2(a)=c_2 a^n \exp (-a^2/\sigma ^2)$ and $w=-1$ and $n=5$ it follows analogously that
\begin{equation}
\label{rhom2}
\rho_{m} = \rho_{m_0}a^{-3}  + K_{2}a^{-3}\left[a^{5}\exp(-a^{2}/\sigma^{2}) - \exp(-1/\sigma^{2})\right]\ ,
\end{equation}
and
\begin{equation}
\rho_{x} = \rho_{x_{0}}^{eff}
- K_{2}\,\exp\left(-a^{2}/\sigma^{2}\right) \left(a^{2} - \frac{3}{2}\sigma^{2}\right)
\ ,
\label{rh2}
\end{equation}
where $K_2\equiv c_2\tilde\rho_{m_0}$ and
\begin{equation}
\rho_{x_{0}}^{eff}
= \rho_{x_{0}} - \frac{3}{2}K_{2}\,\exp (-1/\sigma^{2})
\left[\sigma^{2} - \frac{2}{3}\right]\ .
\label{reffh2}
\end{equation}
Again, in the interaction-free limit $K_{2}\rightarrow 0$ we have $\rho_{x}\rightarrow \rho_{x_{0}} =$ const. The quantity $\rho_{x_{0}}^{eff}$ can be seen as an effective cosmological constant which is re-normalized compared with the ``bare" value, corresponding to $\rho_{x_{0}}$, due to the presence of an interaction.
The ratio $\frac{\ddot{a}}{a}$ becomes
\begin{equation}
\label{accOmega2}
\frac{\ddot a}{a}=-\frac{1}{2}H_{0}^{2}
\left\{\frac{\Omega_{m_{0}} + \Omega_{b_0} - \bar{K}_{2}\exp (-1/\sigma^{2})}{a^3}  -2\Omega_{x_0}^{eff} + 3 \bar{K}_2 \exp
(-a^{2}/\sigma^{2}) \left[a^{2} - \sigma^{2}\right]\right\}
\end{equation}
with $\bar{K}_{2} = \frac{8 \pi G}{3 H_{0}^{2}}K_{2}$ and $\Omega_{x_0}^{eff} = \frac{8 \pi G}{3 H_{0}^{2}}\rho_{x_0}^{eff}$.

\subsection{Decelerated and accelerated expansion}

\noindent  To have a viable cosmological model, formulas (\ref{accOmega1}) and (\ref{accOmega2}) should admit a transition from $\frac{\ddot a}{a} <0$ to $\frac{\ddot a}{a} >0$ before the present time, i.e., for $a < 1$. If, moreover, the accelerated expansion is a transient phenomenon, there should be a change back  from $\frac{\ddot a}{a} >0$ to $\frac{\ddot a}{a} <0$ at a future time, i.e., for $a > 1$.
In both the expressions (\ref{accOmega1}) and (\ref{accOmega2}) the $a^{-3}$ terms on the right hand sides dominate for small values of $a$, i.e., there is decelerated expansion for $a\ll 1$ provided the conditions
\begin{equation}
\label{dec1}
\Omega_{m_{0}} + \Omega_{b_0} > \bar{K}_{1}\exp (-1/\sigma)
\end{equation}
or
\begin{equation}
\label{dec2}
\Omega_{m_{0}} + \Omega_{b_0} > \bar{K}_{2}\exp (-1/\sigma^{2})
\end{equation}
are satisfied.
These conditions put upper limits on the interaction strength. In the non-interacting limit they just express the positivity of the total matter energy density.

Let's consider now the case $a\gg 1$. The dominating contributions in the braces on the right-hand sides  of (\ref{accOmega1}) and
(\ref{accOmega2}) are then  given by the constant terms $-2\rho_{x_0}^{eff}$.
\textit{As long as $\rho_{x_0}^{eff} > 0$ we will have $\frac{\ddot a}{a} >0$ for $a\gg 1$,
i.e., there is no transition back to decelerated expansion}.
This holds, in particular, in the non-interacting limit which reproduces the $\Lambda$CDM model. Then
$\Omega_{x_0}^{eff}$ reduces to $\Omega_{x_0}$, equivalent to $\Omega_{\Lambda_0}$. For $a\gg 1$ this term will always dominate the dynamics.
An obvious way to obtain decelerated expansion for $a\gg 1$ is to
put $\rho_{x_0}^{eff} = 0$ both in (\ref{accOmega1}) and
(\ref{accOmega2}). This corresponds to a vanishing total cosmological constant.
In other words, part of the interaction cancels the ``bare" cosmological constant, described by $\rho_{x_0}$.
Under this condition it is exclusively the remaining part of the interaction which potentially can trigger a period of accelerated expansion.
In such a case one  obtains for the exponential interaction from (\ref{rh1})
\begin{equation}
\Omega_{x_{0}} = 3\bar{K}_{1}\,\exp (-1/\sigma)
\left[\sigma^{2} + \sigma - \frac{1}{3}\right] ,
\label{r10o}
\end{equation}
while from (\ref{rh2}) for the Gaussian interaction
\begin{equation}
\Omega_{x_{0}} = \bar{K}_{2}\,\exp (-1/\sigma^{2})
\left[\frac{3}{2}\sigma^{2} - 1\right]\
\label{r20o}
\end{equation}
is valid.
Adding up the energy contributions $\rho_{b}$ from (\ref{ba}), $\rho_{m}$ from (\ref{rhom1}) and
$\rho_{x }$ with $\rho_{x_0}^{eff}=0$ from (\ref{rh1}), we find for the Hubble function
\begin{equation}
\label{H1}
\frac{H^{2}}{H_{0}^{2}} = \frac{1 - 3 \bar{K}_{1}\exp
(-1/\sigma)\left(\sigma^{2} + \sigma\right)}{a^{3}} + 3 \bar{K}_1 \exp
(-a/\sigma) \left(\sigma^{2} + \sigma a\right)
\ .
\end{equation}
With  $\rho_{b}$ from (\ref{ba}), $\rho_{m}$ from (\ref{rhom2}) and
$\rho_{x }$ with $\rho_{x_0}^{eff}=0$ from (\ref{rh2}), the corresponding expression for the second interaction is
\begin{equation}
\label{H2}
\frac{H^{2}}{H_{0}^{2}} = \frac{1 - \frac{3}{2} \sigma^{2}\bar{K}_{2}\exp
(-1/\sigma^{2})}{a^{3}} + \frac{3}{2} \sigma^{2}\bar{K}_2 \exp
(-a^{2}/\sigma^{2})
\ .
\end{equation}
The acceleration equations (\ref{accOmega1}) and
(\ref{accOmega2}) simplify to
\begin{equation}
\label{accOmega1+}
\frac{\ddot a}{a}=-\frac{1}{2}H_{0}^{2}
\left\{\frac{\Omega_{m_{0}} + \Omega_{b_0} - \bar{K}_{1}\exp (-1/\sigma)}{a^3}  + 3 \bar{K}_1 \exp
(-a/\sigma) \left[a^{2} - 2\sigma^{2} - 2\sigma a\right]\right\}
\end{equation}
and
\begin{equation}
\label{accOmega2+}
\frac{\ddot a}{a}=-\frac{1}{2}H_{0}^{2}
\left\{\frac{\Omega_{m_{0}} + \Omega_{b_0} - \bar{K}_{2}\exp (-1/\sigma^{2})}{a^3} + 3 \bar{K}_2 \exp
(-a^{2}/\sigma^{2}) \left[a^{2} - \sigma^{2}\right]\right\} \ ,
\end{equation}
respectively.
Moreover, with $\Omega_{m_{0}} + \Omega_{b_0} = 1 - \Omega_{x_{0}}$ with $\Omega_{x_{0}}$ from (\ref{r10o}) or (\ref{r20o}), their final forms are
\begin{equation}
\label{accOmega1fin}
\frac{\ddot a}{a}=-\frac{1}{2}H_{0}^{2}
\left\{\frac{1 - 3\bar{K}_1\exp
(-1/\sigma)\left[\sigma^{2} + \sigma\right]}{a^{3}}
+ 3 \bar{K}_1 \exp
(-a/\sigma) \left[a^{2} - 2\sigma^{2} - 2\sigma a\right]\right\}
\end{equation}
and
\begin{equation}
\label{accOmega2fin}
\frac{\ddot a}{a}=-\frac{1}{2}H_{0}^{2}
\left\{\frac{1 - \frac{3}{2}\bar{K}_2\sigma^{2}\exp
(-1/\sigma^{2})}{a^{3}}
+ 3 \bar{K}_2 \exp
(-a^{2}/\sigma^{2}) \left[a^{2} - \sigma^{2}\right]\right\} \ ,
\end{equation}
respectively. To have decelerated expansion for $a\ll 1$,
\begin{equation}
\label{condll1}
 \bar{K}_1 \exp
(-1/\sigma) \left[\sigma^{2} + \sigma\right] < \frac{1}{3}
\end{equation}
or
\begin{equation}
\label{condll2}
 \bar{K}_2 \,\sigma^{2}\,\exp
(-1/\sigma^{2}) < \frac{2}{3}
\end{equation}
has to be required. These conditions are equivalent to (\ref{dec1}) and (\ref{dec2}), respectively.
The zeros of (\ref{accOmega1fin}) and (\ref{accOmega2fin}) determine the values $a_{q}$ of $a$ at which transitions between decelerated and accelerated expansion (or the reverse) occur, namely
\begin{equation}
3 \bar{K}_{1}\,\exp (-1/\sigma)\left[\sigma^{2} + \sigma\right] + 3 \bar{K}_{1}\,
a_{q}^3\,\exp
(-a_{q}/\sigma) \left[2\sigma^{2} + 2\sigma a_{q} - a_{q}^2\right] = 1
\
\label{consist1}
\end{equation}
and
\begin{equation}
\frac{3}{2}\sigma^{2} \bar{K}_{2}\,\exp (-1/\sigma^{2}) +
3 \bar{K}_{2} a_{q}^3\,\exp
(-a_{q}^2/\sigma^2) \left[\sigma^{2} - a_{q}^2\right]
 = 1
\ ,
\label{consist2}
\end{equation}
respectively.
The conditions to have acceleration at the present epoch with $a=1$ are
\begin{equation}
\frac{\ddot{a}}{aH^{2}}\mid _{0} \ > 0 \quad \Leftrightarrow \quad
\bar{K}_{1}\,\exp (-1/\sigma)\left[\sigma^{2} + \sigma - \frac{1}{3}\right] > \frac{1}{9} \
\label{condac1b}
\end{equation}
and
\begin{equation}
\frac{\ddot{a}}{aH^{2}}\mid _{0} \ > 0 \quad \Leftrightarrow \quad
\bar{K}_{2}\,\exp (-1/\sigma^{2})\left[\sigma^{2} - \frac{2}{3}\right] > \frac{2}{9} \ .
\label{condacc1}
\end{equation}
If the inequalities (\ref{condac1b}) or (\ref{condacc1}) hold, we may have present acceleration under the condition $\rho_{x_0}^{eff}=0$, i.e., a vanishing total cosmological constant.
Obviously, the normalized interaction strengths $\bar{K}_{1}$ or  $\bar{K}_{2}$ have to be larger than a threshold value to realize this configuration. The conditions (\ref{condac1b}) or (\ref{condacc1}) are consistent with (\ref{q0x}) if the latter is combined with either  (\ref{r10o}) or (\ref{r20o}), respectively.
On the other hand, we have the upper limits (\ref{condll1}) and (\ref{condll2}). This means, there exists a range for admissible values of the interaction strength, determined by
\begin{equation}
\frac{e^{1/\sigma}}{9\left(\sigma^{2} + \sigma - \frac{1}{3}\right)} < \bar{K}_{1} < \frac{e^{1/\sigma}}{3\left(\sigma^{2} + \sigma\right)}
\label{region1}
\end{equation}
for the first case and
\begin{equation}
\frac{2}{9}\frac{e^{1/\sigma^{2}}}{\sigma^{2} - \frac{2}{3}} < \bar{K}_2 <
\frac{2 e^{1/\sigma^{2}}}{3\sigma^{2}}
\label{region2}
\end{equation}
for the second case.

It is instructive, to compare the relations (\ref{accOmega1fin}) and (\ref{accOmega2fin}) with the corresponding expression
\begin{equation}
\label{acclcdm}
\frac{\ddot a}{a}=-\frac{1}{2}H_{0}^{2}
\left\{\frac{1 - \Omega_{\Lambda}}{a^{3}}
- 2 \Omega_{\Lambda}\right\} \
\end{equation}
of the $\Lambda$CDM model.
Obviously, the interaction terms
in (\ref{accOmega1fin}) and (\ref{accOmega2fin}) may play the role of $\Omega_{\Lambda}$ in (\ref{acclcdm}).
Now we know, that the $\Lambda$CDM model provides a fairly good description of the present universe, i.e., for $a=1$. Therefore, we do not expect a radically different scenario around the present epoch, based on any alternative model. This suggest positive values of the interaction constants $\bar{K}_{1}$ and $\bar{K}_{2}$ together with $\sigma >1$. It will turn out that these models are indeed favored.

\subsection{Direction of the energy transfer and structure of the source terms}

From (\ref{Q2}) it is obvious, that in an expanding universe ($\dot{a} > 0$) the quantity $\frac{d g}{d a}$ should be positive to have an energy transfer from dark energy to dark matter.
Positive values of $Q$ are preferred on thermodynamical grounds \cite{pavon}. (But see \cite{schaefer,alcaniz09}.)
For the case $g_{1} = c_{1}a^{n}\exp (-a/\sigma)$ we have
\begin{equation}
\frac{d g_{1}}{d a} = \left(\frac{n}{a} - \frac{1}{\sigma}\right) g_{1}\ .
\label{dg1a}
\end{equation}
It follows that for $n=5$ and for $c_{1}>0$, equivalent to $\bar{K}_{1}>0$,
\begin{equation}
a < 5 \sigma \quad\Leftrightarrow \quad Q>0 \ .
\label{Q1>}
\end{equation}
For $g_{2} = c_{2}a^{n}\exp (-a^{2}/\sigma^{2})$
\begin{equation}
\frac{d g_{2}}{d a} = \left(\frac{n}{a} - \frac{2a}{\sigma^{2}}\right) g_{2}
\label{dg2a}
\end{equation}
is valid.
This means, for $n=5$ and for $c_{2}>0$, equivalent to $\bar{K}_{2}>0$,
\begin{equation}
a^{2} < \frac{5}{2} \sigma^{2} \quad\Leftrightarrow \quad Q>0 \ .
\label{Q2>}
\end{equation}
Consequently, for sufficiently large values of the parameter $\sigma$, positive interaction constants $\bar{K}_{1}$ and $\bar{K}_{2}$ are both thermodynamically favored and provide a description of the present cosmological epoch which is similar to the well-tested $\Lambda$CDM model (see the comment following Eq.~(\ref{acclcdm})). With $w=-1$ our setting is reminiscent of models with a decaying cosmological ``constant". Notice that for future values $a > 1$ for which the conditions (\ref{Q1>}) or (\ref{Q2>}) are violated, the interaction is no longer operative.

The explicit forms of the interaction term $Q$ in the balances (\ref{interacao}) become
\begin{equation}
Q_{1} = K_{1}a^{2}\,H\,\left(5 - \frac{a}{\sigma}\right)\exp{\left(-a/\sigma\right)} \
\label{Q1a}
\end{equation}
for the first interaction and
\begin{equation}
Q_{2} = K_{2}a^{2}\,H\,\left(5 - 2\frac{a^{2}}{\sigma^{2}}\right)\exp{\left(-a^{2}/\sigma^{2}\right)} \
\label{Q2a}
\end{equation}
for the second one. These terms may be written as
\begin{equation}
Q_{1} = \beta_{1}\left(a\right) H \rho_{x}
\label{Q1rho}
\end{equation}
with
\begin{equation}
\beta_{1}\left(a\right) = \frac{1}{3} \frac{5 - \frac{a}{\sigma}}
{\frac{\sigma^{2}}{a^{2}} + \frac{\sigma}{a} - \frac{1}{3}}
\
\label{b1}
\end{equation}
for the first interaction and as
\begin{equation}
Q_{2} = \beta_{2}\left(a\right) H \rho_{x}
\label{Q2rho}
\end{equation}
with
\begin{equation}
\beta_{2}\left(a\right) = \frac{2}{3} \frac{5 - 2\frac{a^{2}}{\sigma^{2}}}
{\frac{\sigma^{2}}{a^{2}} - \frac{2}{3}}
\
\label{b2}
\end{equation}
respectively, for the second interaction.
Interactions of the type $Q = \beta H \rho_{x}$ with an interaction constant $\beta$ have been frequently used to describe interactions between dark matter and dark energy \cite{WDCQG,Bin,Maartens}. Here we have generalized this structure to a variable interaction parameter. For both the cases (\ref{b1}) and (\ref{b2}) this parameter vanishes for small values of the scale factor, i.e., the interaction is switched on during the cosmological evolution.
In the late time limit both $Q_{1}$ and  $Q_{2}$ vanish because $\rho_{x}$ itself vanishes.

\subsection{Scalar field representation}

In this subsection we point out  that there also exists an equivalent scalar field representation for the dynamics of the interacting
$x$ component. Rewriting the balance equation for the $x$ component (cf. Eq.~(\ref{interacao})) as
\begin{equation}
\dot{\rho_x}+3H(1+w_{eff})\rho_x = 0
\  \label{ebxeff}
\end{equation}
with
\begin{equation}
w_{eff} = w + \frac{Q_{1}}{3H\rho_x}
\  \label{weff}
\end{equation}
for the first interaction and introducing the correspondence
\begin{equation}
\left(1 + w_{eff}\right)\rho_x = \dot{\phi}^{2}
\ , \label{dphi}
\end{equation}
we find
\begin{equation}
\dot{\phi}^{2}=  \frac{1}{3} K_1 a^2 \left(5 - \frac{a}{\sigma}\right)\exp (-a/\sigma) \
\label{dphi21}
\end{equation}
and, via
\begin{equation}
V = \frac{1}{2}\frac{1 - w_{eff}}{1 + w_{eff}}\,\dot{\phi}^{2}
\ ,
\label{Vdef}
\end{equation}
an effective potential
\begin{equation}
V = 3 K_{1}a^{2}\left[\frac{\sigma^{2}}{a^{2}} + \frac{\sigma}{a} - \frac{11}{18} + \frac{a}{18\sigma}\right]
\,\exp (-a/\sigma)
\ .
\label{V1(a)}
\end{equation}
The kinetic part vanishes for very small and very large values of the scale factor. The potential starts with a constant initial value $V_{i} \approx 3 K_{1}\sigma^{2}$ at $a\ll 1$ and also tends to zero for $a\gg 1$.
Eq.~(\ref{dphi21}) is equivalent to
\begin{equation}
\frac{d\phi}{da} =  \frac{\left[\frac{\bar{K}_{1}}{8\pi G}
\left(5 - \frac{a}{\sigma}\right)\exp\left(-a/\sigma\right)\right]^{1/2}}{\frac{H}{H_{0}}}
\label{dphi1da}
\end{equation}
with $\frac{H}{H_{0}}$ from (\ref{H1}). The solution $\phi\left(a\right)$ of (\ref{dphi1da}) together with (\ref{V1(a)}) yields an implicit representation of $V(\phi)$.

The corresponding relations for the second interaction are
\begin{equation}
\dot{\phi}^{2}=  \frac{1}{3} K_2 a^2 \left(5 - 2\frac{a^{2}}{\sigma^{2}}\right)\exp (-a^{2}/\sigma^{2}) \
\label{dphi22}
\end{equation}
and
\begin{equation}
V = \frac{3}{2} K_{2}a^{2}
\left[\frac{\sigma^{2}}{a^{2}} + \frac{2}{9} \frac{a^{2}}{\sigma^{2}} - \frac{11}{9} \right]
\,\exp (-a^{2}/\sigma^{2})
\
\label{V2(a)}
\end{equation}
with a similar behavior as in the previous case.
Eq.~(\ref{dphi22}) may be written
\begin{equation}
\frac{d\phi}{da} =  \frac{\left[\frac{\bar{K}_{2}}{8\pi G}
\left(5 - 2\frac{a^{2}}{\sigma^{2}}\right)\exp\left(-a^{2}/\sigma^{2}\right)\right]^{1/2}}{\frac{H}{H_{0}}}\ ,
\label{dphi2da}
\end{equation}
now with $\frac{H}{H_{0}}$  from (\ref{H2}).

\begin{center}
\begin{figure}[!t]
\begin{minipage}[t]{0.4\linewidth}
\includegraphics[width=\linewidth]{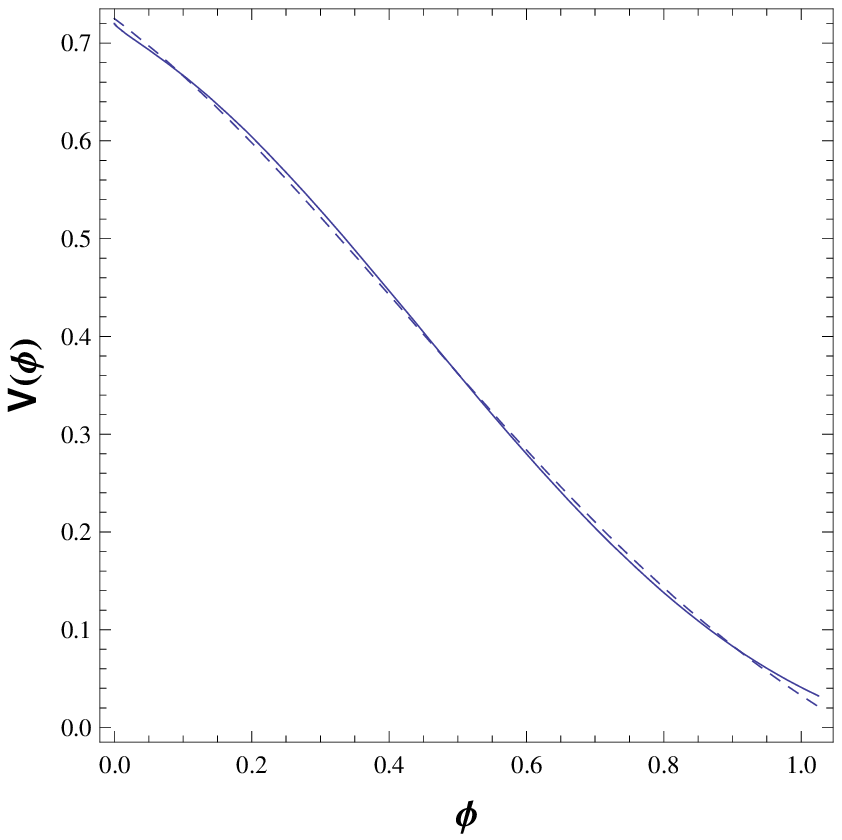}
\end{minipage} \hfill
\begin{minipage}[t]{0.4\linewidth}
\includegraphics[width=\linewidth]{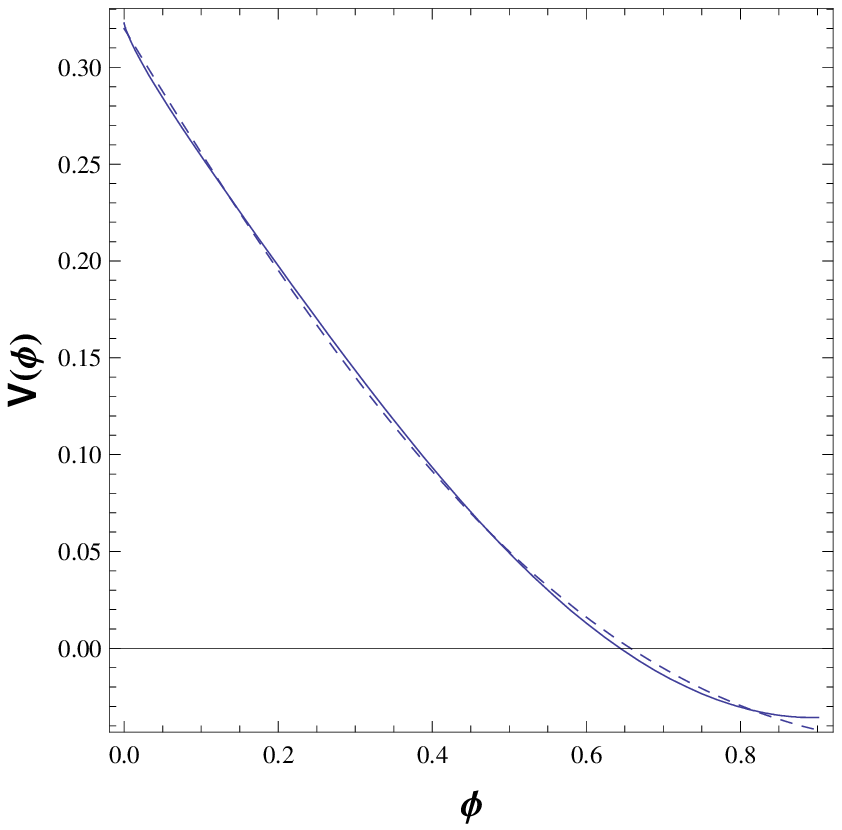}
\end{minipage} \hfill
\caption{{\label{potential}\protect\footnotesize Left panel: effective potential of  
the scalar field representation for the first interaction, right panel: effective potential of
the scalar field representation for the second interaction.  
The continuous lines represent the numerical results, the dashed lines the interpolation in
terms of power and $\cosh$ functions.}}
\end{figure}
\end{center}

Both potentials can be determined, as function of $\phi$, numerically. The potentials can approximately be expressed by a finite
series in terms of power and $\cosh$ functions. For example, the series can have the form 
\begin{equation}
V(\phi) = a_0 + \frac{a_1\phi}{\cosh\phi} + \frac{a_2}{\cosh^2\phi}.
\end{equation}
For the first interaction we find $a_0 = 0.121151$, $a_1 = - 0.527741$ and $a_2 = 0.603898$, while for the second interaction we have $a_0 = 0.416204$, $a_1 = -0.655256$ and $a_2 = -0.0962396 $. Figure (\ref{potential}) displays the numerical interpolation and the approximate analytical
expression for each of the cases.

\section{Supernova Type Ia constraints}
\label{SN}

To constraint the free parameters of the model, we use the gold sample of the SNIa data \cite{gold}. The crucial quantity to
be evaluated is the luminosity distance
\begin{equation}
D_L = c\,(1 + z)\int_0^z\frac{dz'}{H\left(z'\right)}\ ,
\end{equation}
where $H$ is given either by (\ref{H1}) or by (\ref{H2}). The quantity $H_0 = 100h\,\mathrm{km/Mpc/s}$ in these expressions is the Hubble parameter today.
The observational data are expressed in terms of the moduli distance
\begin{equation}
\mu_0 = \log\biggr(\frac{D_L}{Mpc}\biggl) + 25\ .
\end{equation}
From this quantity the statistical function $\chi^2$ is constructed. It is defined
by
\begin{equation}
\chi^2 = \sum_i\frac{(\mu_{0i}^t - \mu_{0i}^o)}{\sigma_i^2}\ ,
\end{equation}
where $\mu_{0i}^t$ is the predicted theoretical value for the moduli distance for the $i^{th}$ supernova, obtained by using the model developed above, $\mu_{0i}^o$ is the corresponding observational data, $\sigma_i$ being its observational error bar.
With the help of $\chi^2$ we construct the probability density function (PDF)
\begin{equation}
P(x_n) = A\,e^{-\chi^2(x_n)/2}\ ,
\end{equation}
where $x_{n}$ denotes the set of free parameters of the model; $A$ is a normalization constant.
The general description of the Bayesian analysis can be found in reference \cite{sn1}.
\par
\begin{center}
\begin{figure}[!t]
\begin{minipage}[t]{0.25\linewidth}
\includegraphics[width=\linewidth]{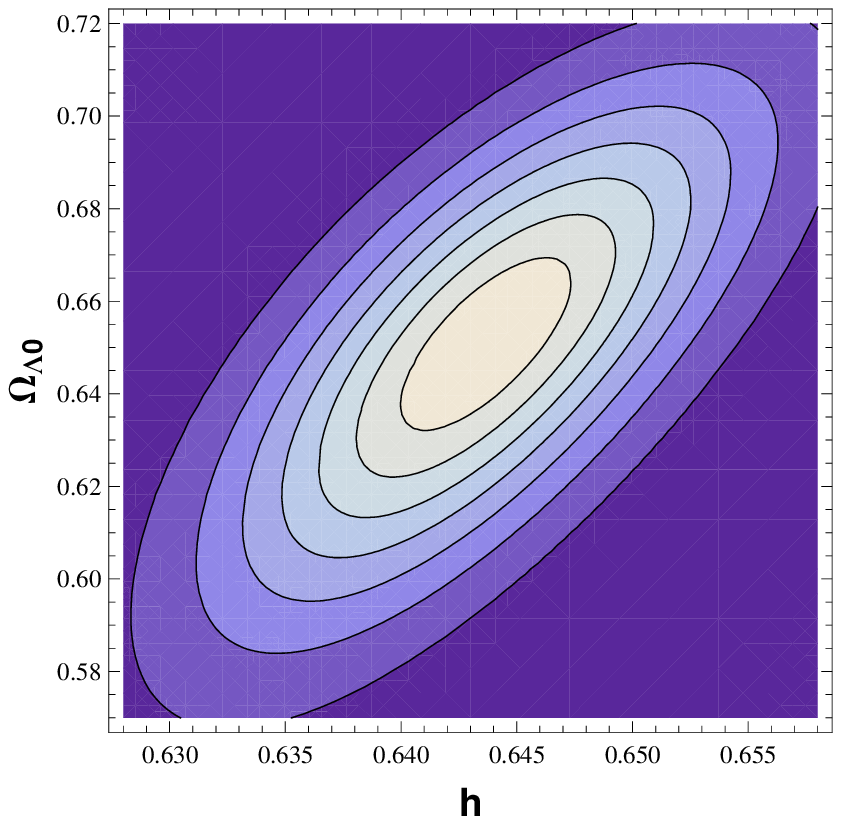}
\end{minipage} \hfill
\begin{minipage}[t]{0.25\linewidth}
\includegraphics[width=\linewidth]{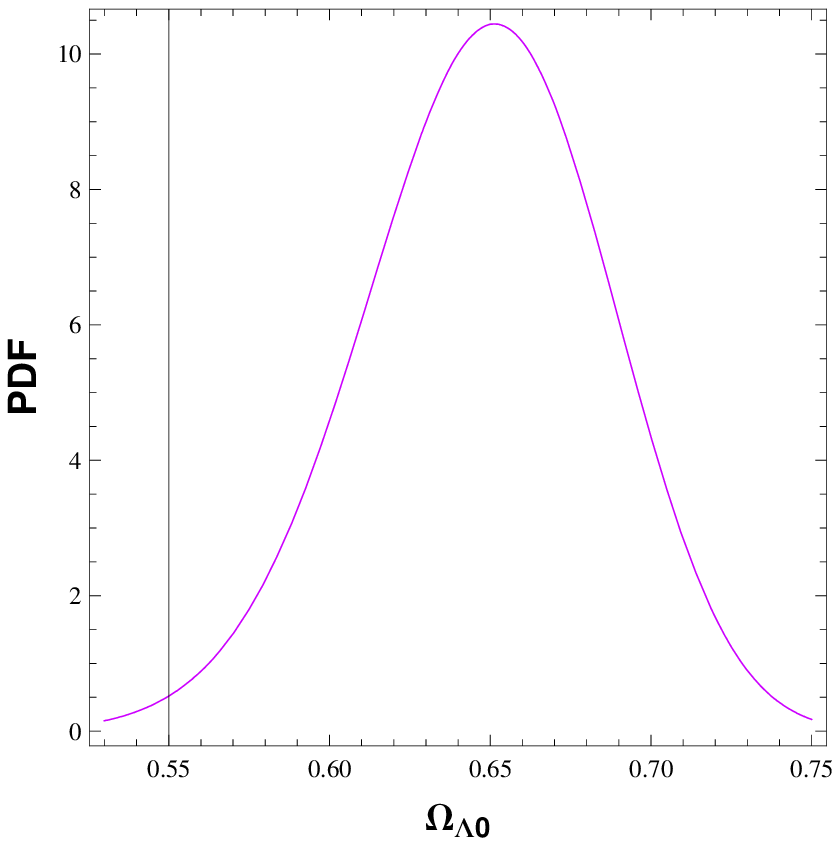}
\end{minipage} \hfill
\begin{minipage}[t]{0.25\linewidth}
\includegraphics[width=\linewidth]{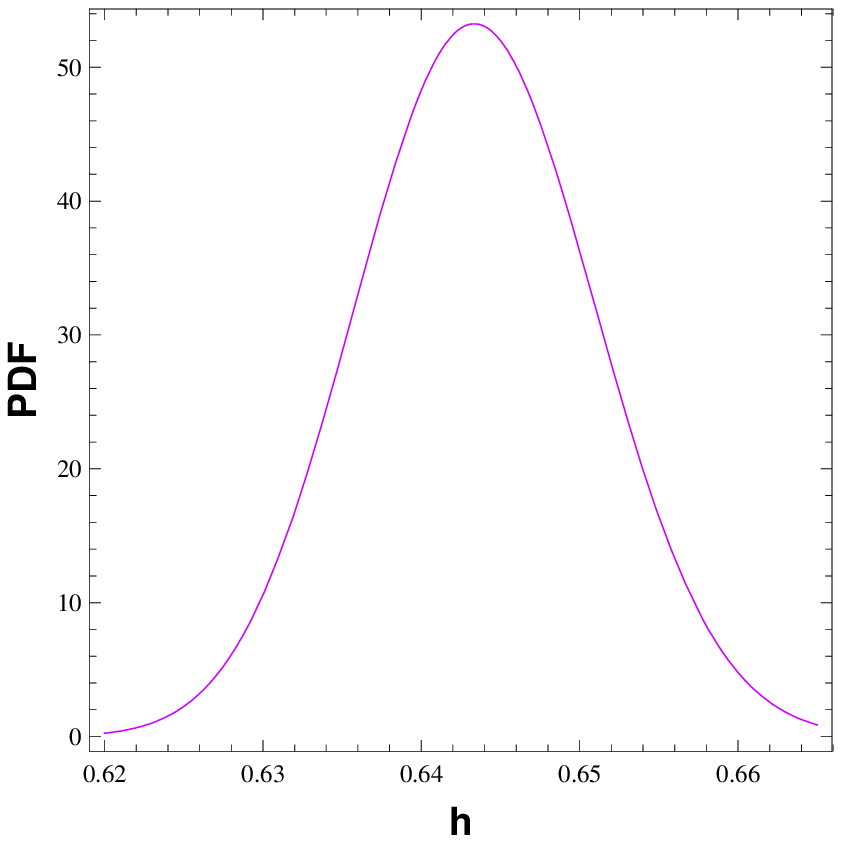}
\end{minipage} \hfill
\caption{{\label{lcdm}\protect\footnotesize The $\Lambda$CDM model. Left panel: the two-dimensional probability distribution for $h$ and $\Omega_{\Lambda0}$. The brighter the color, the higher the probability.  Center panel: one-dimensional probability distribution function for $\Omega_{\Lambda0}$. Right panel: one-dimensional probability distribution function for $h$.}}
\end{figure}
\end{center}

At first we obtain the results for the $\Lambda$CDM model which is given by the non-interacting limits $\bar{K}_{1} = 0$ and $\bar{K}_{2} = 0$ for the specific configurations considered here. The $\Lambda$CDM model has only two free parameters: the density parameter $\Omega_{x0} = \Omega_{\Lambda0}$
and the Hubble parameter $h$. Minimizing $\chi^2$, we find $\chi^2_{min} = 1.128$ with
$\Omega_{x0} = \Omega_{\Lambda0} = 0.651$ and $h = 0.644$. The two-dimensional PDF for these parameters is shown  in figure \ref{lcdm} (left). After marginalization,
we find the one-dimensional PDFs for $\Omega_{\Lambda0}$ and $h$, also displayed in figure \ref{lcdm} (center and right, respectively). After
marginalization, the maximum probabilities occur for $\Omega_{\Lambda0} = 0.651$ and $h = 0.619$.

In the  interacting model there are generally four free parameters to be constrained: $\Omega_{x0}$, $\bar K_1$ or $\bar K_2$, the Hubble parameter $h$ and $\sigma$. But our transient scenario implies the relations (\ref{r10o}) or (\ref{r20o}) between $\Omega_{x0}$ and $\bar{K}_1$ or $\bar{K}_2$, respectively.
This reduces
the number of free parameter to three: $\bar{K}_1$ or $\bar{K}_2$, $h$ and $\sigma$. Integrating over one (two) of them we can obtain the two-dimensional (one-dimensional) PDFs for the remaining variable(s).
From now on, we will restrict ourselves to the three-dimensional
phase space case.

For the first interaction (interaction strength parameter $\bar{K}_{1}$) the relevant equations are (\ref{H1}) and (\ref{accOmega1fin}). The corresponding relations for the second interaction (interaction strength parameter $\bar{K}_{2}$) are (\ref{H2}) and (\ref{accOmega2fin}).

\begin{center}
\begin{figure}[!t]
\begin{minipage}[t]{0.35\linewidth}
\includegraphics[width=\linewidth]{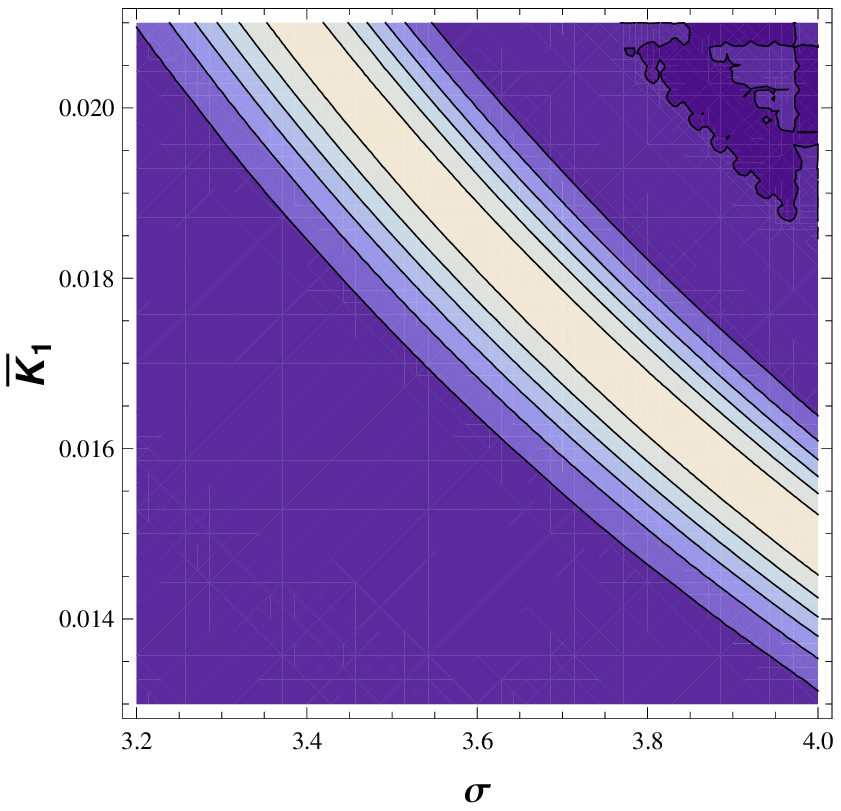}
\end{minipage} \hfill
\begin{minipage}[t]{0.35\linewidth}
\includegraphics[width=\linewidth]{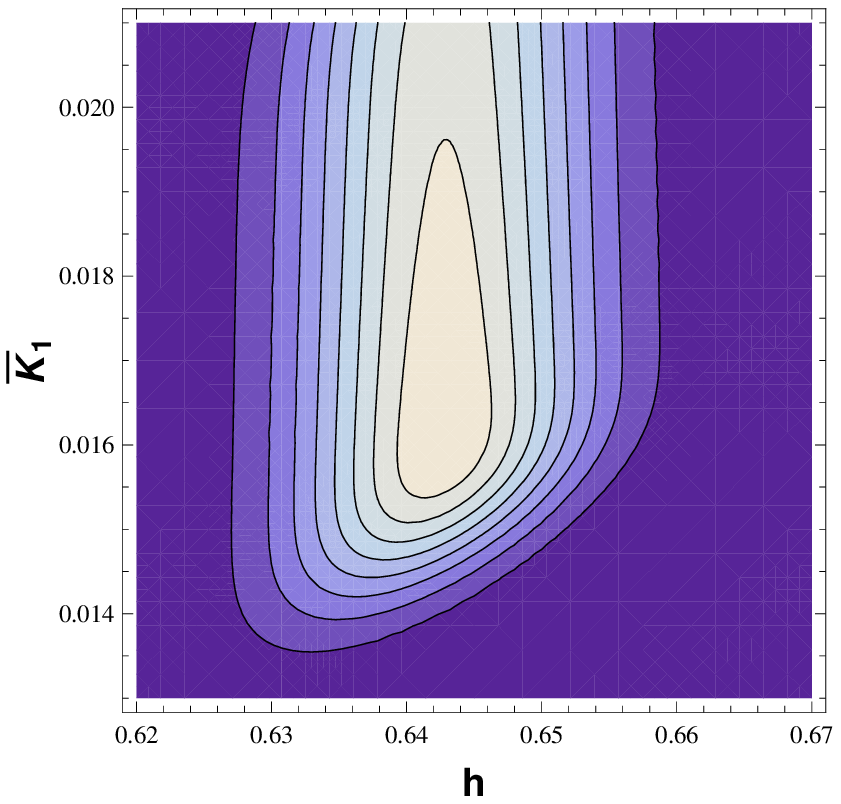}
\end{minipage} \hfill
\begin{minipage}[t]{0.35\linewidth}
\includegraphics[width=\linewidth]{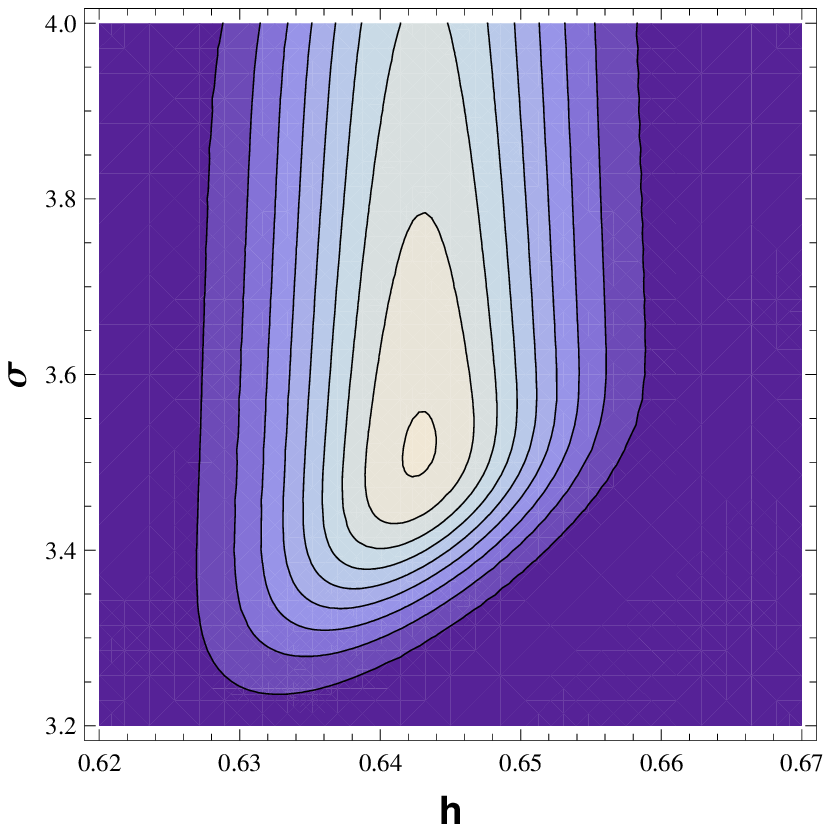}
\end{minipage} \hfill
\begin{minipage}[t]{0.35\linewidth}
\includegraphics[width=\linewidth]{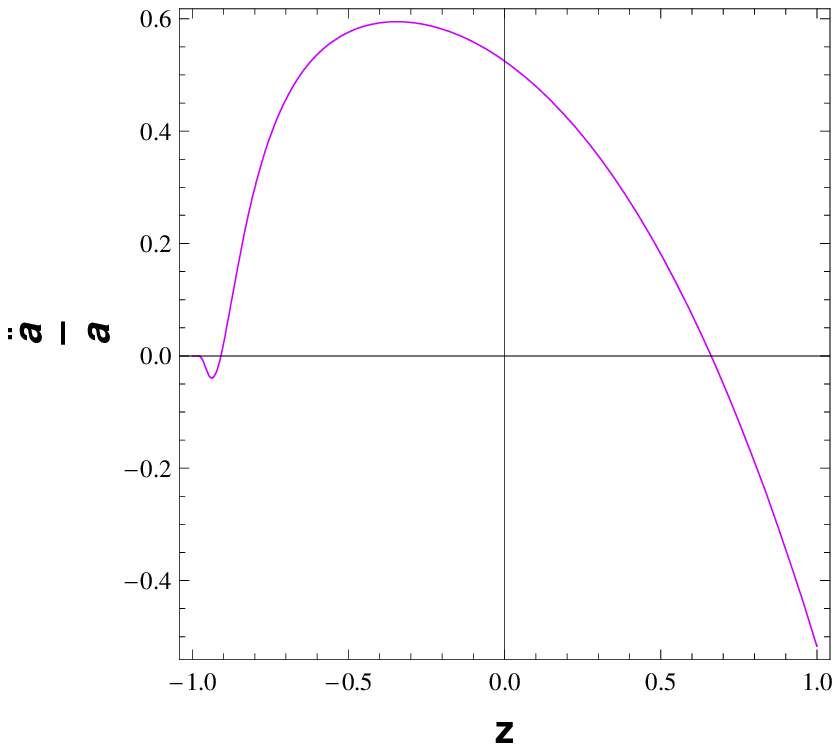}
\end{minipage} \hfill
\caption{\label{2dbis}{\protect\footnotesize Two-dimensional probability distributions for
different combinations of $\bar K_1$,
$\sigma$ and $h$ after marginalization over the remaining parameter. The lower right panel shows the evolution of the acceleration parameter for the best fit scenario
in units of $H_{0}^{2}$.}}
\end{figure}
\end{center}

\begin{center}
\begin{figure}[!t]
\begin{minipage}[t]{0.35\linewidth}
\includegraphics[width=\linewidth]{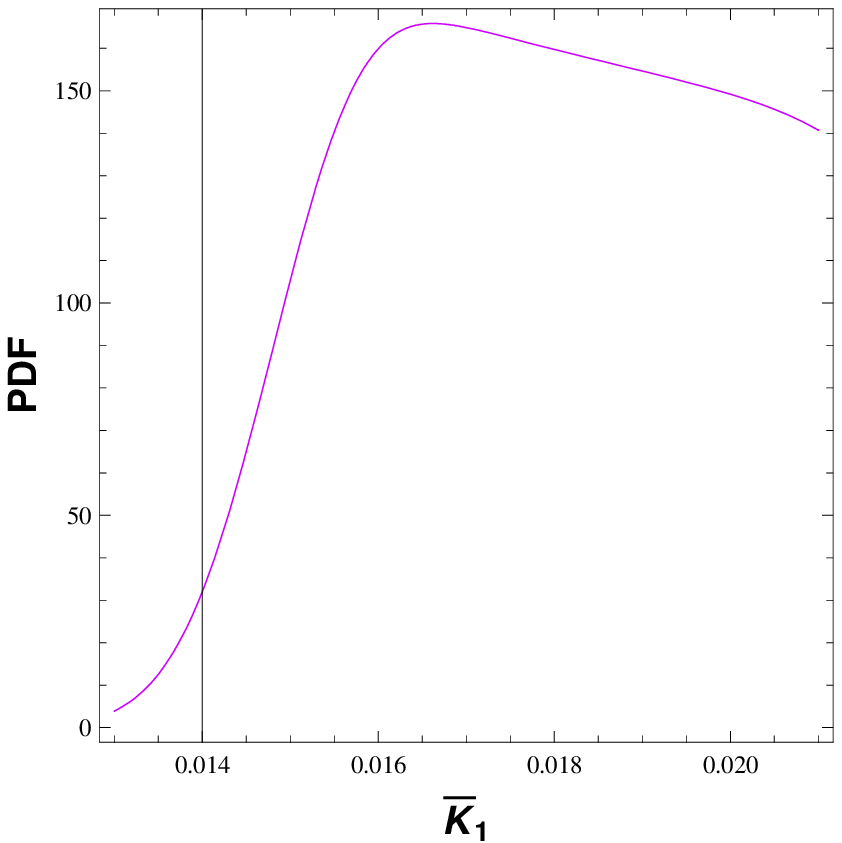}
\end{minipage} \hfill
\begin{minipage}[t]{0.35\linewidth}
\includegraphics[width=\linewidth]{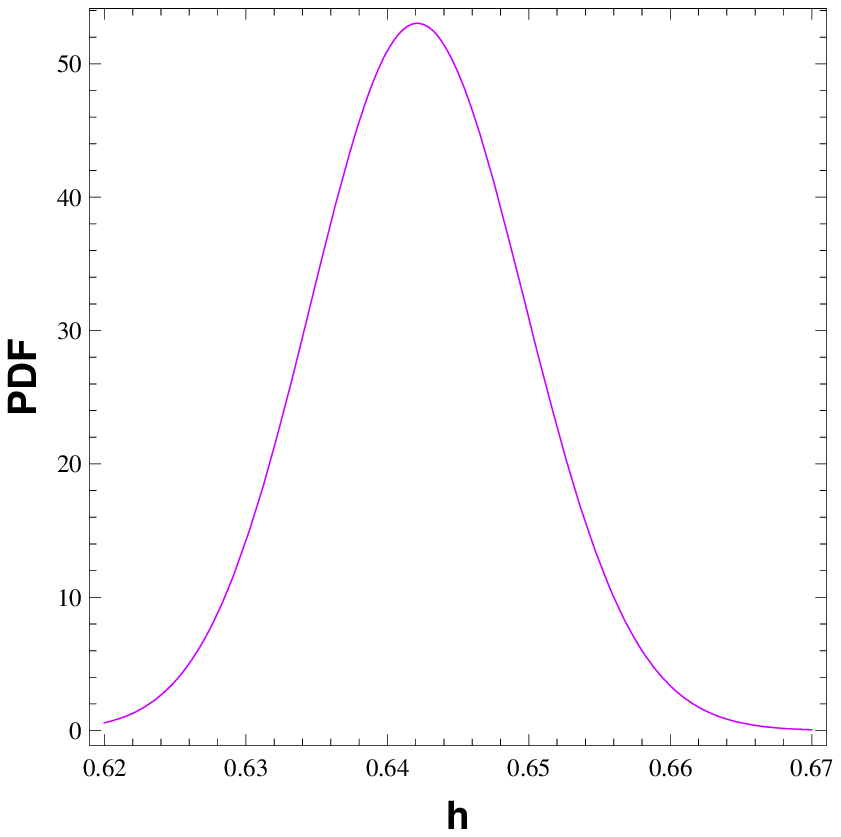}
\end{minipage} \hfill
\begin{minipage}[t]{0.35\linewidth}
\includegraphics[width=\linewidth]{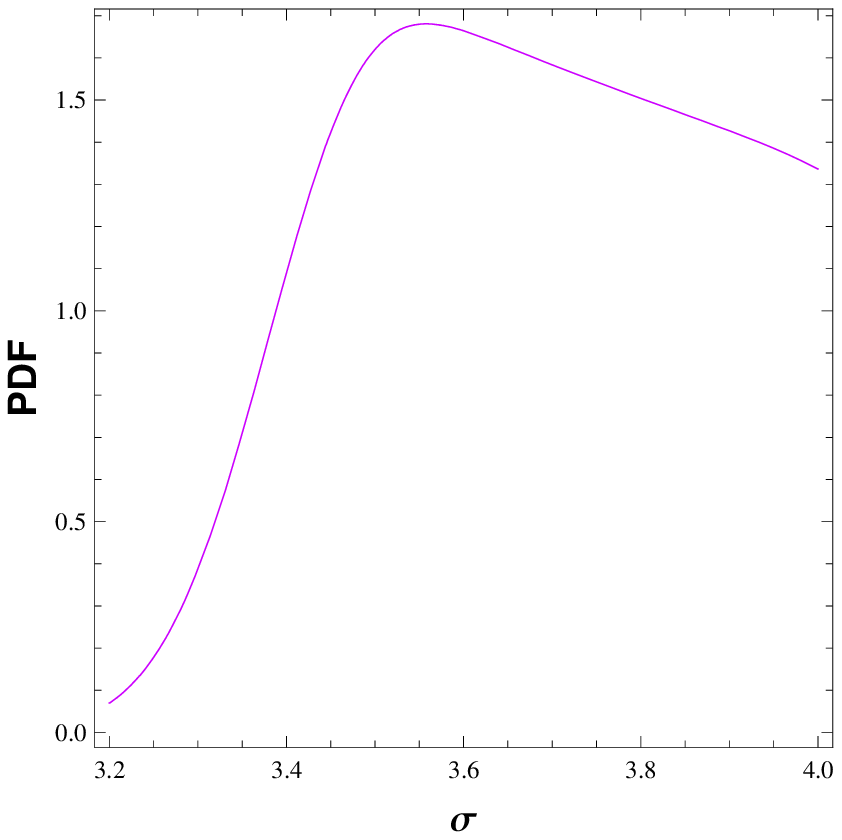}
\end{minipage} \hfill
\begin{minipage}[t]{0.35\linewidth}
\includegraphics[width=\linewidth]{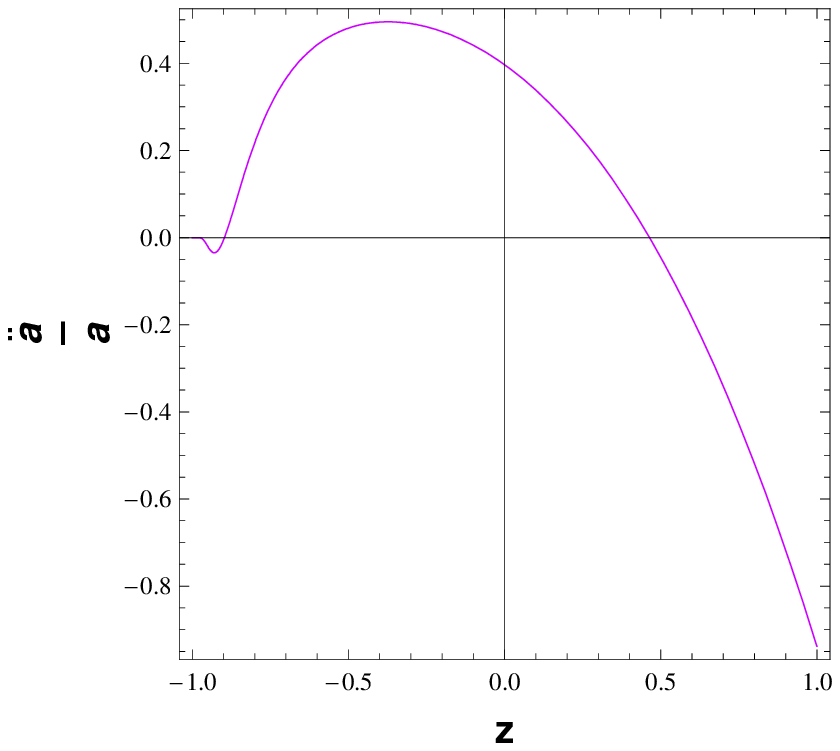}
\end{minipage} \hfill
\caption{\label{1dbis}{\protect\footnotesize One-dimensional probability distributions for
$\bar K_1$, $\sigma$ and $h$ after marginalization over the two remaining parameters. The lower right panel shows the evolution of the acceleration parameter for the highest peak probability values in units of $H_{0}^{2}$.
}}
\end{figure}
\end{center}

For the first interaction we obtain the two-dimensional and one-dimensional PDFs
shown in figures \ref{2dbis} and \ref{1dbis}. The best fit is obtained
for $h = 0.643$, $\sigma = 4.000$ and $\bar K_1 = 0.015$ with $\chi^2_{min} = 1.129$, corresponding to $\Omega_{x0} = 0.683$. The evolution
of the acceleration parameter for this set of values is shown in the lower right panel of figure \ref{2dbis}. It is obvious that after an intermediate accelerated expansion the Universe reenters a decelerating phase.
After marginalization, the peaks of the one-dimensional PDFs occur for $h = 0.670$, $\sigma = 3.558$ and $\bar K_1 = 0.017$. These values
imply $\Omega_{x0} = 0.599$ and again a temporary period of accelerated expansion.
The corresponding
evolution of the acceleration parameter is displayed in the lower right panel of figure \ref{1dbis}.

\begin{center}
\begin{figure}[!t]
\begin{minipage}[t]{0.35\linewidth}
\includegraphics[width=\linewidth]{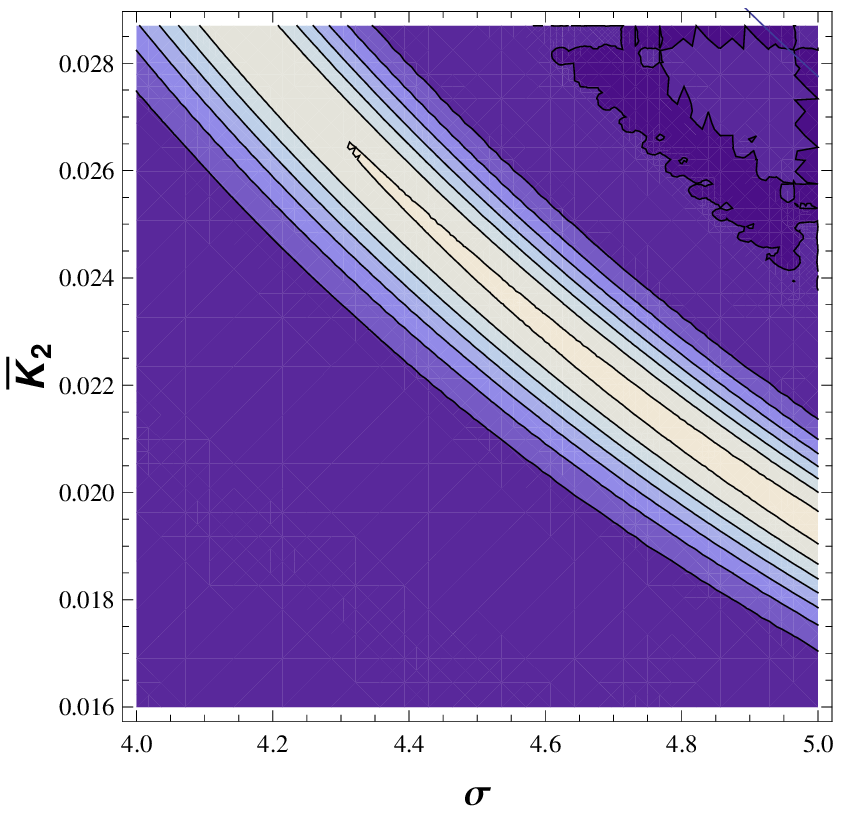}
\end{minipage} \hfill
\begin{minipage}[t]{0.35\linewidth}
\includegraphics[width=\linewidth]{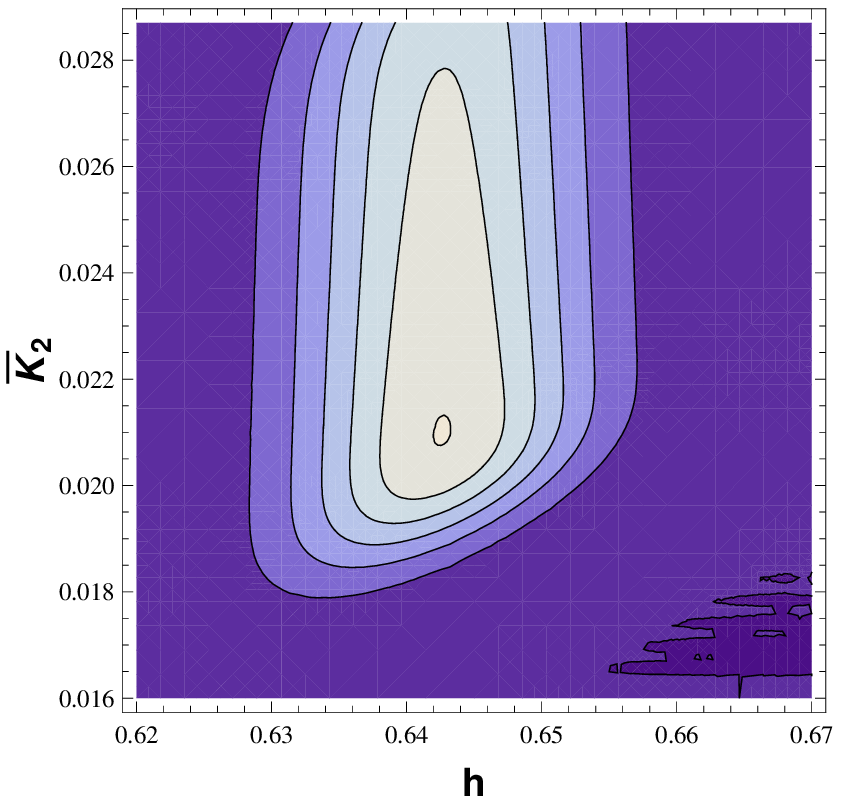}
\end{minipage} \hfill
\begin{minipage}[t]{0.35\linewidth}
\includegraphics[width=\linewidth]{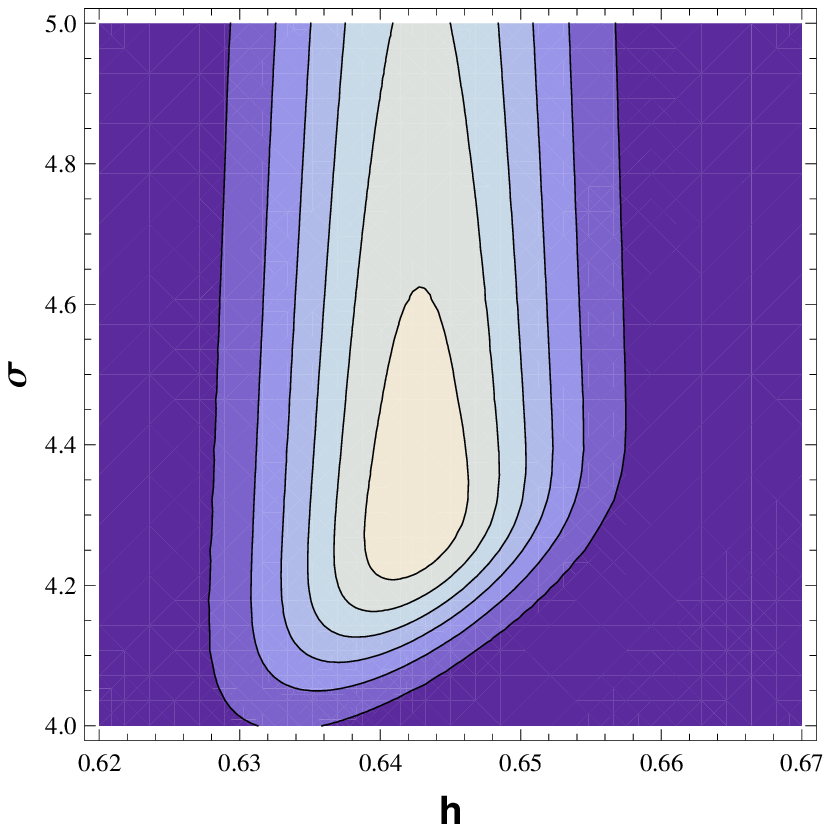}
\end{minipage} \hfill
\begin{minipage}[t]{0.35\linewidth}
\includegraphics[width=\linewidth]{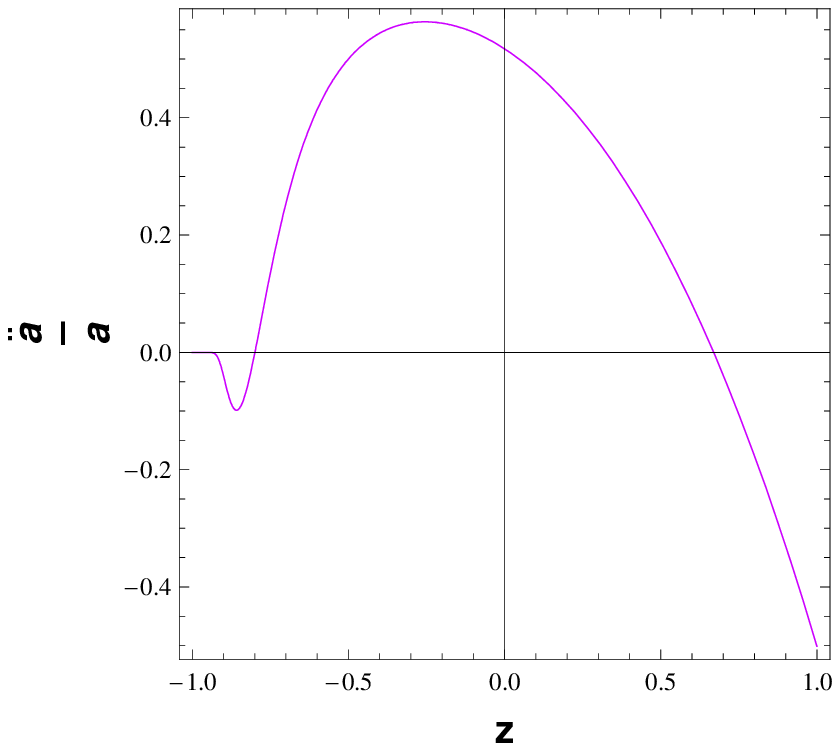}
\end{minipage} \hfill
\caption{\label{2d}{\protect\footnotesize Two-dimensional probability distributions for the different combinations
of $\bar K_2$, $\sigma$ and $h$ after marginalization over the remaining parameter. The lower right panel shows the evolution of the acceleration parameter for the best fit scenario
in units of $H_{0}^{2}$.}}
\end{figure}
\end{center}

\begin{center}
\begin{figure}[!t]
\begin{minipage}[t]{0.35\linewidth}
\includegraphics[width=\linewidth]{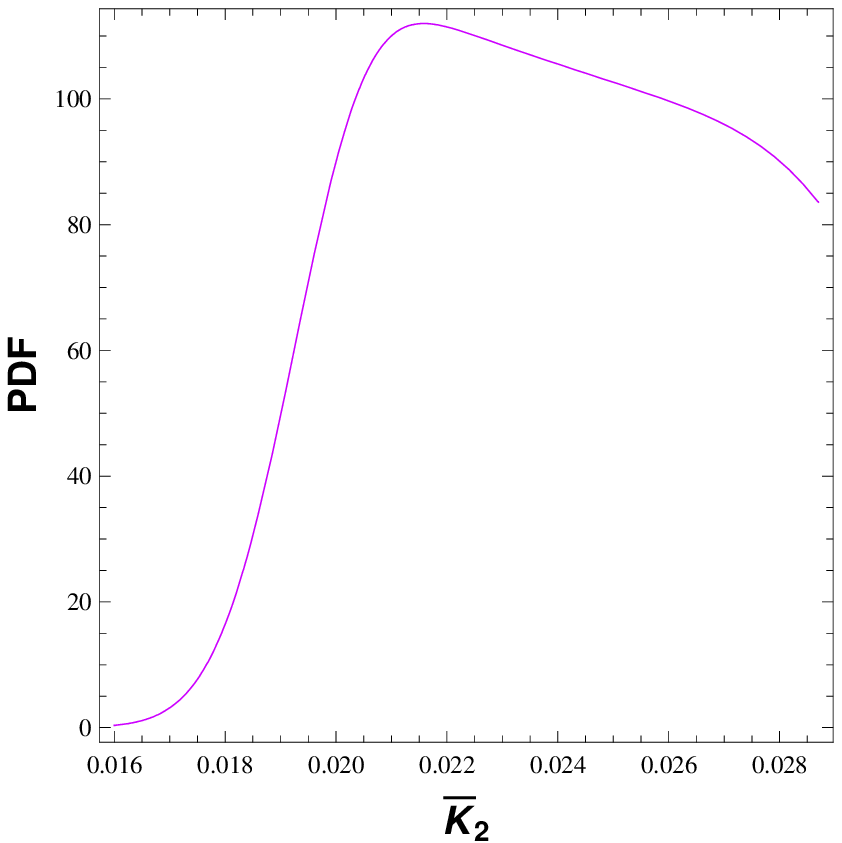}
\end{minipage} \hfill
\begin{minipage}[t]{0.35\linewidth}
\includegraphics[width=\linewidth]{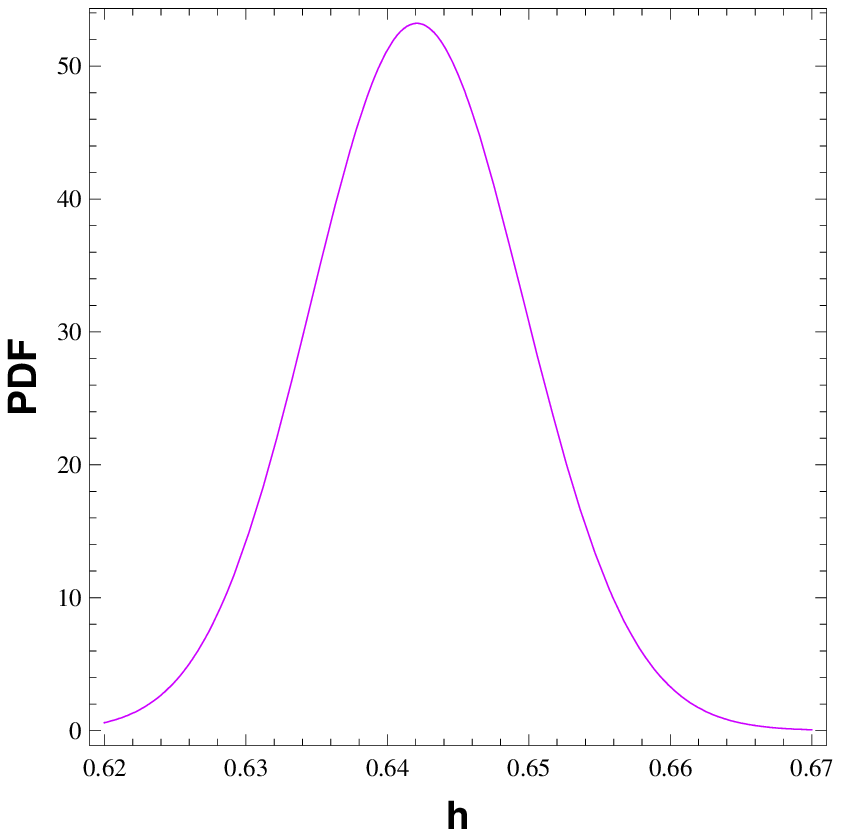}
\end{minipage} \hfill
\begin{minipage}[t]{0.35\linewidth}
\includegraphics[width=\linewidth]{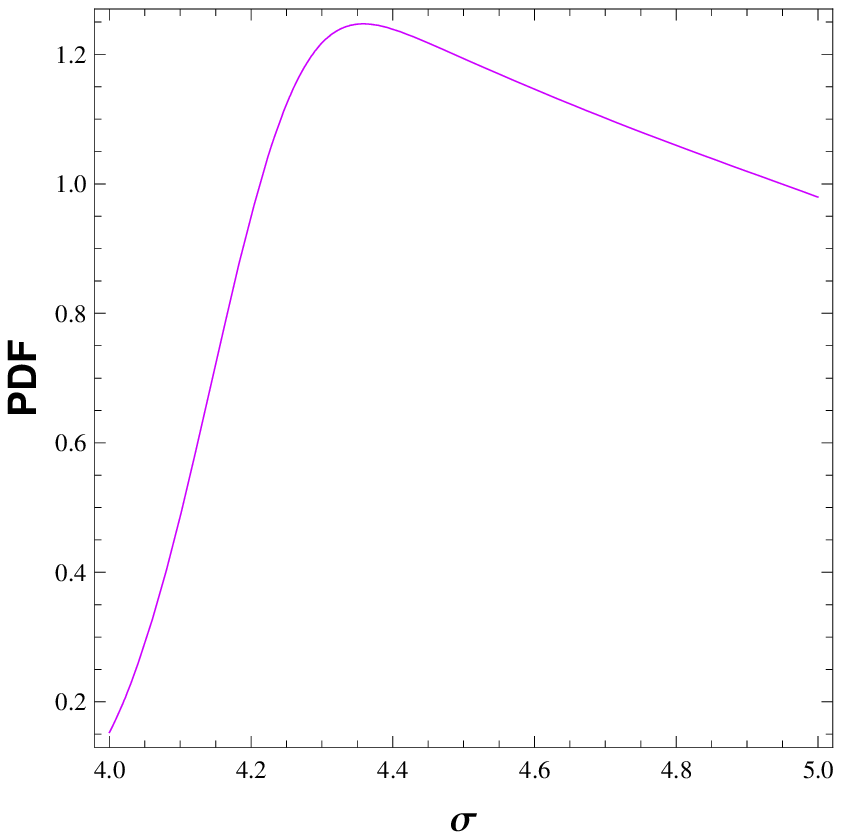}
\end{minipage} \hfill
\begin{minipage}[t]{0.35\linewidth}
\includegraphics[width=\linewidth]{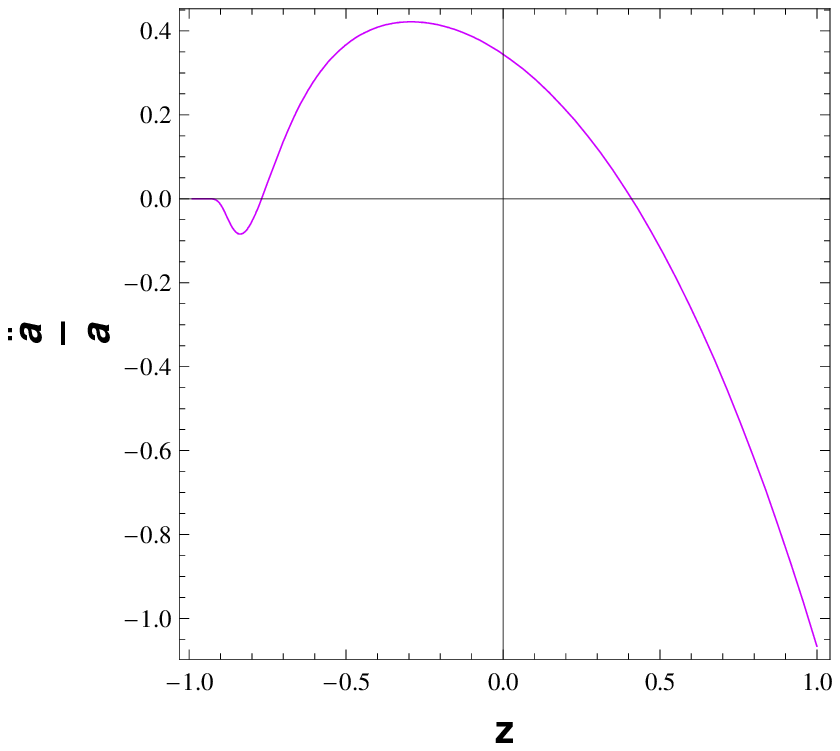}
\end{minipage} \hfill
\caption{\label{1d}{\protect\footnotesize  One-dimensional probability distributions for
$\bar K_2$, $\sigma$ and $h$  after marginalization over the two remaining parameters. The lower right panel shows the evolution of the acceleration parameter for the highest peak probability values in units of $H_{0}^{2}$.
}}
\end{figure}
\end{center}

The results for the second interaction are very similar. In this case we find the two-dimensional and one-dimensional PDFs
shown in figures \ref{2d} and \ref{1d}. The best-fit values are $h = 0.644$, $\sigma = 5.000$ and $\bar K_2 = 0.019$, with $\chi^2_{min} = 1.129$, corresponding to $\Omega_{x0} = 0.678$. The evolution
of the acceleration parameter is shown in the lower right panel of figure \ref{2d}.
After marginalization, the peaks of the one-dimensional PDFs are located at $h = 0.670$, $\sigma = 4.358$ and $\bar K_2 = 0.022$, implying $\Omega_{x0} = 0.563$.
The
behavior of the acceleration parameter for this set of values is displayed in the lower right panel of figure \ref{1d}, confirming again that the accelerated expansion of the universe is a transient phenomenon.
We recall that all the cases of figures \ref{2dbis} - \ref{1d} correspond to an energy transfer from dark energy to dark matter.



\section{Conclusions}
\label{conclusions}

In a homogeneous and isotropic cosmological background, a fluid with an equation of state $w=-1$ is dynamically equivalent to a cosmological constant. According to the $\Lambda$CDM model, a small (in Planck units), but so far theoretically unexplained, cosmological constant is responsible for the observed accelerated expansion of our present Universe.
We have demonstrated that an interaction between such type of dark energy and dark matter re-normalizes
this constant. Even if the resulting effective cosmological constant is assumed to be zero, the time-dependent part of the interaction (the part that does not contribute to the re-normalization) was shown to be able to generate a phase of accelerated expansion of the Universe. We have studied two types of interactions for which cosmic acceleration is a transient phenomenon. While recent investigations favor a scenario in which the acceleration of the Universe is already slowing down today \cite{sastaro}, our toy model predicts the maximal acceleration to occur at a future time. 
The $\Lambda$CDM model is unable to provide any type of transient acceleration dynamics. A statistical analysis of the SNIa data shows that models of transient accelerated expansion may well be competitive with the standard cosmological constant model.
While the mentioned cancelation mechanism of the ``bare" cosmological constant is purely phenomenological, we believe that it might possibly indicate a more fundamental feature. A vanishing total effective cosmological constant seems more appealing and easier to understand within an underlying basic theory than a small, non-vanishing value of this quantity.

In general, our conclusion concerning the potentially transient character of the cosmic acceleration as the result of an interaction in the dark sector coincides with the numerical results of \cite{alcaniztr}. But we think that an analytic solution, even though it is the solution of a toy model, provides additional information about the specific role of the interaction in this scenario and thus gives rise to a physically more transparent picture.

{\bf Acknowledgement}  We thank FAPES and CNPq (Brazil) for
financial support.


%
\end{document}